\documentclass[cits]{PoS}

\newcommand{\hiir}{H~{\scshape ii}~region}
\newcommand{\hiirs}{H~{\scshape ii}~regions}
\newcommand{\hiiR}{H~{\scshape ii}~Region}
\newcommand{\um}{$\mu$m}

\title{Beyond Str\"{o}mgren Spheres and Wind-Blown Bubbles: An Observational Perspective on \hiiR~Feedback}

\ShortTitle{Bash Symposium '11---H~{\footnotesize II} Region Feedback}

\author{\speaker{Matthew S. Povich}%
        \thanks{NSF Astronomy and Astrophysics Postdoctoral Fellow.}\\
       Penn State\\
       E-mail: \email{povich@astro.psu.edu}}

\abstract{Massive stars produce copious quantities of ultraviolet radiation beyond the Lyman limit, photoionizing the interstellar medium (ISM) and producing \hiirs. As strong sources of recombination- and forbidden-line emission, infrared continuum, and thermal (free-free) radio continuum, \hiirs\ serve as readily-observable beacons of massive star formation in the Milky Way and external galaxies. Along with supernovae, \hiirs\ are dominant sources of feedback in star-forming galaxies, injecting radiative and mechanical luminosity into the ISM. \hiirs\ may prove more important than supernovae as triggers of star formation through localized compression of cold cloud cores. In this review, I give a broad overview of the structure and time-evolution of \hiirs, emphasizing complications to the theoretical picture revealed by multiwavelength observations. I discuss a recent controversy surrounding the dominant feedback mechanism in 30 Doradus, the most luminous \hiir\ in the Local Group. I summarize the first results from the Milky Way Project (MWP), which has produced a new catalog of several thousand candidate Galactic \hiirs\ by enlisting >35,000 "citizen scientists" to search {\it Spitzer Space Telescope} survey images for bubble-shaped structures. The MWP and similar large catalogs enable empirical studies of Galactic \hiir\ evolution across the full range of luminosities and statistical studies of triggered star formation.
}

\FullConference{Frank N. Bash Symposium 2011: New Horizons in Astronomy\\
		 October 9-11, 2011 \\
		 Austin, Texas, USA}

\begin{document}

\section{Introduction: The Interstellar Medium Out of Balance}

Most students of astronomy will encounter a course on the physics of the interstellar medium (ISM; the gas and dust occupying the space between the stars) at some point in their undergraduate or post-graduate studies. Hence the concept that the ISM exists in several distinct ``phases,'' in rough pressure equilibrium with each other, may be familiar to most readers. Physical quantities characterizing the major phases of ISM gas in the Milky Way are summarized in Table~\ref{tab1}, compiled from the textbooks by Tielens \cite{Tielens 2005} and Draine \cite{Draine 2011}. The coexistence of gas at such dramatically different densities, temperatures, and ionization fractions in the same galaxy is explained by the different filling factors of each phase. While the precise values for the filling factors remain poorly-measured and controversial, the basic picture that the colder, denser phases of the ISM exists as smaller clouds within the more diffuse, warm/hot phases, is well established. \hiirs, localized regions of photo-ionized gas produced by hot, massive, OB-type stars, occupy a negligible fraction of the ISM volume, hence perhaps \hiirs\ ought not to be regarded as a proper ISM phase at all. However, massive stars form in the densest regions of cold, molecular clouds, and as their far-ultraviolet (UV) radiation first photo-dissociates molecules and then photo-ionizes atoms, the multi-phase physics of the ISM can be studied within a single, small volume. 

\begin{table}[hb] 
\begin{center}
  \begin{tabular}{lccccc}
  \hline\hline
    Phase & Density & $T$ & Total Mass & Scaleheight & Filling \\
          & (cm$^{-3}$) &  (K) & ($10^9$ M$_{\odot}$) & (pc) & factor \\
    \hline
    Hot ionized medium & ${\sim}0.004$ & ${\sim}10^6$ & --- & 3000 & ${\sim}0.5$ \\
    Warm neutral medium & 0.5--0.6 & 8000 & 2.8 & ${\sim}300$ & ${\sim}0.4$ \\
    Warm ionized medium & 0.1--0.3 & ${\sim}5000$ & 1.0 & 900 & ${\sim}0.1$ \\
    Cold neutral medium & 30--50 & 80--100 & 2.2 & 100 & ${\sim}0.01$ \\
    Molecular Clouds & $10^2$--$10^6$ & 10--50 & 1.3 & 75 & ${\sim}10^{-4}$ \\
    \hline
    \hiiR s & 1--$10^4$ & $10^4$ & 0.05 & 70 & --- \\
  \hline 
\end{tabular} \caption{Phases of ISM Gas in the Milky Way} 
  \label{tab1}
\end{center} 
\end{table}

In one of the immortalized insights of early modern astrophysics, Str{\"o}mgren \cite{Stromgren 1939} realized that the mean free path in neutral hydrogen of UV photons beyond the Lyman limit ($\lambda < 912$~\AA) is negligibly small compared to the size of the ionized hydrogen region produced by a hot star. \hiirs\ therefore have sharp boundaries, or I-fronts, where the recombination rate balances the ionization rate. For a single star in an ambient medium of constant density, this boundary is defined as the Str{\"o}mgren radius,
\begin{equation}
  R_{S0} = \left(\frac{3Q_0}{4\pi n_H^2 a_B}\right)^{1/3},
\end{equation}
where $Q_0$ is the ionizing photon rate (which depends on the star or stars responsible for the \hiir), $n_H$ is the hydrogen gas density, and $a_B$ is the Case B recombination coefficient \cite{Spitzer 1978}. Generally speaking, main-sequence or giant stars earlier than B3 emit sufficient $Q_0$ to produce observable Galactic \hiirs.

Because the ionizing stars provide an internal source of radiative and mechanical luminosity, \hiirs\ rapidly become overpressured compared to the ambient ISM and expand. If the dominant source of pressure is collisional heating of gas by free electrons, then the time-evolution of the expansion follows the simple analytic relation from Lyman Spitzer's classic ISM text \cite{Spitzer 1978}:
\begin{equation}
  R_S(t) = \left(1 + \frac{7}{4}\frac{c_{s2}t}{R_{S0}}\right)^{4/7},
\end{equation}
where $c_{s2}\sim 10$~km~s$^{-1}$ is the sound speed in the ionized gas. The expansion velocity obtained by differentiating this relation can exceed the (significantly lower) sound speed in the ambient medium, hence expanding I-fronts often become shock waves. If the ambient medium is molecular, UV photons emerging from the \hiir\ destroy the molecules, creating a photodissociation region (PDR) around the \hiir.

\begin{figure}[htb] 
\begin{center}
  \includegraphics[width=.7\textwidth]{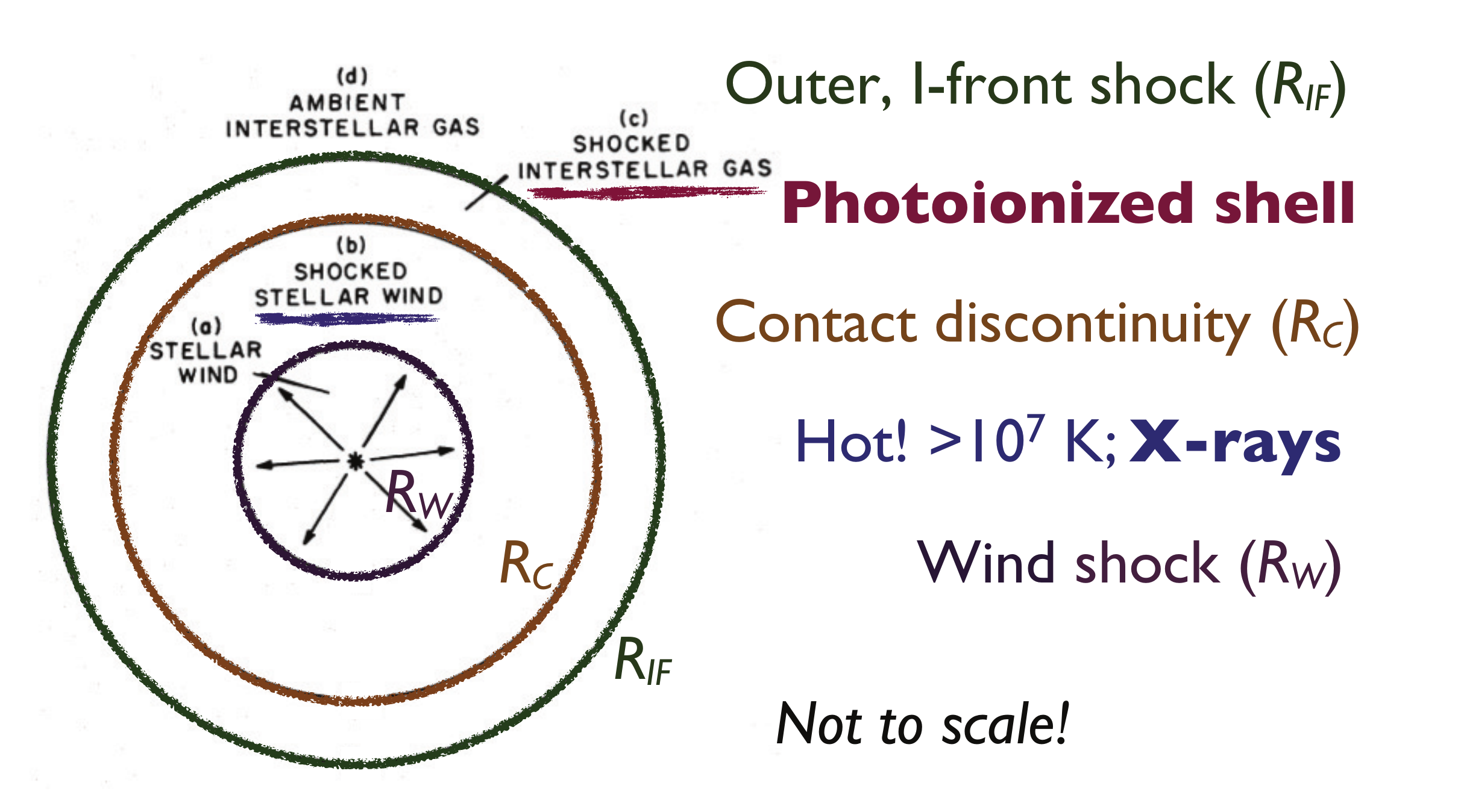} 
  \caption{Anatomy of a wind-blown bubble, adapted from Weaver et
    al.~\cite{Weaver et al. 1977}} \label{WBB} 
\end{center}
\end{figure}

{\it Early} O stars and OB giants drive powerful winds that fundamentally alter the structure of more luminous \hiirs.
Castor et al.\ \cite{Castor et al. 1975} and Weaver et al.\ \cite{Weaver et al. 1977} provided analytical models for wind-blown bubbles produced by isolated, hot stars, and numerous authors have subsequently refined these models using a variety of semi-analytical and numerical techniques \cite{McKee et al. 1984,Koo+McKee 1992,Capriotti+Kozminski 2001,Freyer et al. 2003,Harper-Clark+Murray 2009}. The basic, ``onion-layer'' structure of a wind-blown bubble is illustrated in Figure~\ref{WBB}. The highly supersonic (1000--2000 km s$^{-1}$) stellar wind flows freely outward for a short distance $R_W$ from the star before it is shocked, producing a bubble of very hot, ionized gas. The ``classical'' \hiir\ collapses into a photoionized shell of gas, (imperfectly) separated from the hot gas zone by a contact discontinuity at $R_C$. The I-front at $R_{IF}$ still represents the outer boundary of the \hiir. 

Many complications separate the ideal models of Str{\"o}mgren spheres and wind-blown bubbles from reality. The ISM is clumpy, hence the ambient medium surrounding an \hiir\ is never uniform. Massive stars tend to form in clusters, hence multiple stars often contribute to the ionization of a single \hiir. The ambient ISM is generally in motion with respect to the ionizing star(s), and stellar winds need not be spherically symmetric. Dust mixed with gas in \hiirs\ enables dramatic, radiative cooling \cite{Everett+Churchwell 2010}, and magnetic fields threading through the clouds contribute anisotropic pressure support \cite{Pellegrini et al. 2007}. Turbulence provides additional pressure and facilitates mixing at the interfaces between gas layers. Early models neglected completely the contribution of radiation pressure \cite{Krumholz+Matzner 2009,Draine 2011b}.

\section{Multiwavelength Observations of \hiiR s}

\begin{figure} 
\begin{center}
  \includegraphics[width=.7\textwidth]{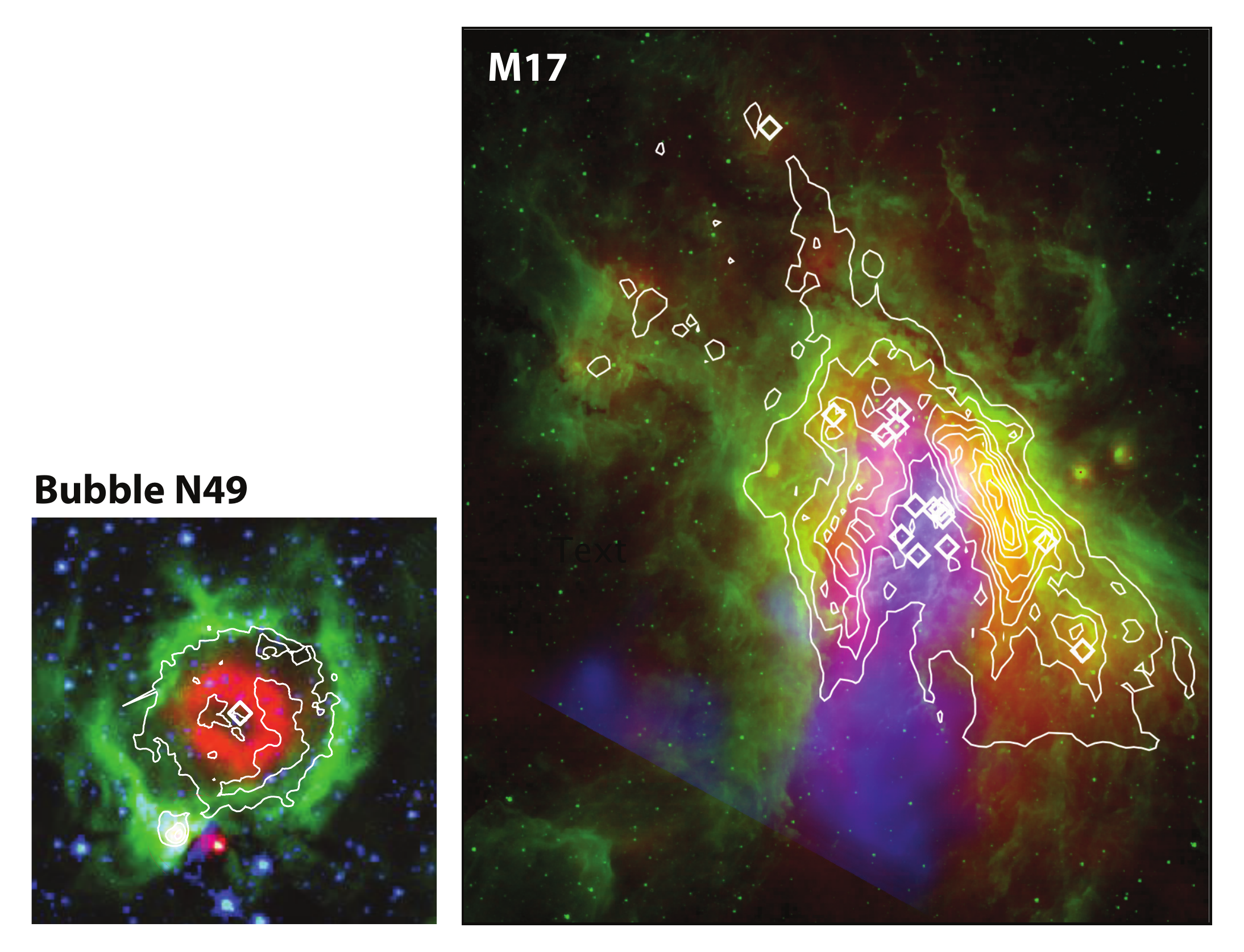} 
  \caption{Multiwavelength images of the wind-blown bubble N49 \cite{Watson et al. 2008} and the
  giant \hiir\ M17 \cite{Povich et al. 2007}, displayed at approximately the same physical scale. N49 image: red = {\it Spitzer}/MIPS 24~\um, green/blue = {\it Spitzer}/IRAC 8.0/4.5~\um. M17 image: red = {\it MSX} 21.3~\um, green = IRAC 5.8~\um, blue = {\it Chandra} soft (0.5--2 keV), diffuse X-rays. In both images, diamonds denote known O and early B stars and contours show 20 cm thermal radio continuum.} \label{N49+M17} 
\end{center}
\end{figure}

In spite of the messy complexity governing the structure of real \hiirs, the basic structures predicted by the wind-blown bubble models are identifiable in modern, multiwavelength images. In Figure~\ref{N49+M17}, a prototypical wind-blown bubble ionized by a single O6.5 V star ($Q_0 = 8.5\times 10^{48}$ s$^{-1}$) \cite{Churchwell et al. 2006, Watson et al. 2008} is compared to the giant \hiir\ M17, ionized by a dozen O stars, including several O4 V stars ($Q_0 = 3\times 10^{50}$ s$^{-1}$) \cite{Povich et al. 2007}. In N49, the I-front at $R_{IF}$ is defined by the sharp inner rim of 8.0~\um\ (green) polycyclic aromatic hydrocarbon (PAH) emission from the PDR, neatly encapsulating the photoionized gas shell (contours). Dust mixed with the photoionized gas and heated by radiation from the central star forms a torus of 24~\um\ emission. Everett \& Churchwell \cite{Everett+Churchwell 2010} found that the lifetime of dust grains within the harsh environment of this \hiir\ is extremely short, and suggested that dust must be continuously replenished from evaporating dense clumps to produce the observed 24~\um\ emission. Draine \cite{Draine 2011b} demonstrated that radiation pressure can produce evacuated cavities in the centers of \hiirs, but noted that the central hole in N49 is too large to be explained by radiation pressure alone and suggested that the stellar wind also contributes. An interface analogous to the contact discontinuity $R_C$ in Figure~\ref{WBB} is likely located within the radio shell/24~\um\ torus in N49, hence $R_C/R_{IF} < 1/2$. 

Unlike N49, M17 is far from round, yet similar morphological features can be discerned, with one important addition (Figure~\ref{N49+M17}). A spectacular plume of hot, X-ray-emitting plasma (blue) occupies the central cavity of M17 \cite{Townsley et al. 2003}. This X-ray emission provides {\it direct} evidence for stellar wind shocks. At an absorption-corrected X-ray luminosity $L_X = 7 \times 10^{34}$~erg~s$^{-1}$, this plasma is unusually bright in comparison to other \hiirs, but {\it it is fainter than the predictions of wind-blown-bubble theory by more than an order of magnitude} \cite{Townsley et al. 2011}. This discrepancy may be explained by collisional interactions with dust grains providing a mechanism for cooling the hot plasma, and/or depressurization of the wind-blown bubble where the plasma is not completely confined by the nebula. Either interpretation implies that the contact discontinuity in Figure~\ref{WBB} does not effectively separate the photoionzed gas/dusty shell from the hot gas bubble in M17. Again assuming $R_C$ falls at the inner edge of the photoionized shell and heated dust emission, $R_C/R_{IF}\approx 1/2$ in M17. The inner cavity in M17 is clearly larger (both in absolute volume and as a fraction of the \hiir\ volume) in comparison to that of N49. The photoionized shell in M17 is supported by a combination of radiation pressure and hot gas pressure \cite{Pellegrini et al. 2007}.

\begin{figure}[b] 
\begin{center}
  \includegraphics[width=.95\textwidth]{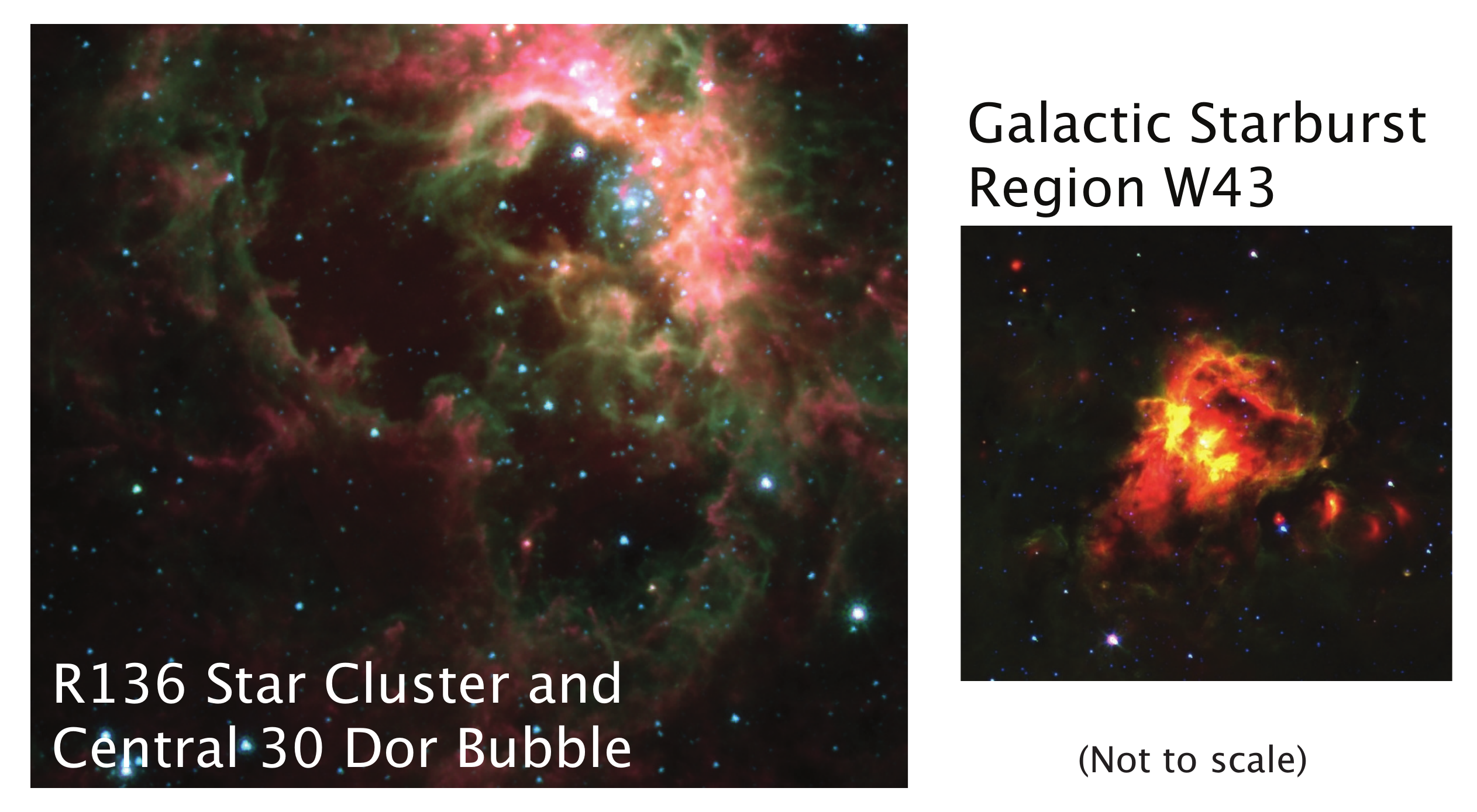} 
  \caption{{\it Spitzer} mid-IR images of two starburst \hiirs. Left: Image of 30 Dor in the Large Magellanic Cloud, red = 8.0 \um, green = 4.5 \um, blue = 3.6 \um. Right:  Image of W43, red = 24~\um, green = 8.0~\um, blue = 4.5 \um. } \label{30Dor+W43} 
\end{center}
\end{figure}

Although N49 and M17 are representative of a range of Galactic \hiirs\ where stellar winds play an important role, on the Galactic scale feedback is dominated by the most luminous, starburst regions. In Figure~\ref{30Dor+W43}, a Galactic starburst region, W43, is compared to 30 Doradus in the Large Magellanic Cloud (LMC), the most luminous \hiir\ in the Local Group ($Q_0=4.2\times 10^{51}$ s$^{-1}$; \cite{Pellegrini et al. 2011}). Mid-infrared (IR) images of the starburst \hiirs\ reveal layered bubble morphologies that are remarkably similar between the two regions. The large bubble lobes shown in each panel of Figure~\ref{30Dor+W43} are part of larger \hiir\ complexes, with the ionizing clusters partially (in 30 Dor) or completely (W43) obscured by dense, foreground filaments of bright, mid-IR emission. In the 30 Dor image, the PDRs appear pink and the photo-ionized shells green, while the W43 image matches the color-code of Figure~\ref{N49+M17}, in which the PDRs appear yellow-green and dust mixed within photoionized shell appears red. Here it is most appropriate to describe the photoionized gas structures as shells, occupying thin layers just interior to the PDRs, with $R_C \approx R_{IF}$. Indeed, the bubbles in 30 Dor are known to be filled with hot, X-ray-emitting plasma \cite{Townsley et al. 2006}, and similar plasma would likely be found in W43 if comparable {\it Chandra X-ray Observatory} observations were obtained.

Comparing Figures~\ref{N49+M17} and \ref{30Dor+W43}, a trend becomes apparent: as the luminosity of the ionizing cluster increases, so does $R_C/R_{IF}$, the size of the central cavity relative to the overall size of the \hiir. We may extend this trend down to low-luminosity \hiirs\ ionized by late O or early B stars, for which the central cavity disappears entirely ($R_C=0$). It will be useful to bear this trend in mind when considering the controversial issue of precisely which feedback mechanism, hot gas pressure or direct radiation pressure, dominates \hiir\ structure at the high-luminosity extreme.

\section{A Rumble in the Tarantula: What is the Dominant Feedback Mechanism in 30 Doradus?}

Thanks to its status as the most luminous \hiir\ in the Local Group and its location in the low-metallicity environment of the LMC, 30 Dor (popularly known as the Tarantula Nebula) has long received intense observational scrutiny. At $d = 50$~kpc, 30 Dor is the best nearby laboratory for studying the physical conditions that prevailed in the unresolvable, high-redshift star-forming regions that dominated the major cosmological epoch of galaxy-building \cite{Pellegrini et al. 2011}. Recently, Lopez et al.\ \cite[hereafter L11]{Lopez et al. 2011} and Pellegrini et al.\ \cite[hereafter P11]{Pellegrini et al. 2011} carried out independent, parallel studies of the feedback processes shaping 30 Dor. Using fundamentally different approaches to interpreting multiwavelength datasets, these authors reached diametrically opposed conclusions; L11 reported that direct stellar radiation pressure dominates the interior of the \hiir, while P11 argued that the pressure of the hot, X-ray-emitting plasma shapes the large-scale structure and dynamics.
This disagreement is rooted in the different definitions of radiation pressure and the different assumed nebular geometries used in the two studies. 

L11 used the simplest definition of direct radiation pressure, 
\begin{equation}\label{L11}
  P_{\rm dir} = \sum{\frac{L_{\rm bol}}{4\pi r^2c}}
\end{equation}
(their equation 1), where $L_{\rm bol}$ is the bolometric luminosity of each star and $r$ is the distance traveled by the starlight to reach a given point in the nebula, deprojected assuming a spherical geometry. 
$P_{\rm dir}$ declines sharply with distance from the central star cluster, R136 (note that this expression diverges for $r=0$). 

By contrast, P11 constructed a non-symmetric, cavity model (based on the central region of 30 Dor shown in Figure \ref{30Dor+W43}) for the nebular geometry and used photoionization models to calculate the density of H atoms $n_H$ and hence the ionization parameter $U$ at each position,
\begin{equation}\label{P11U}
  U = \frac{Q_0}{4\pi r^2 c n_H}.
\end{equation}
The divergent behavior of this expression is avoided by implementing the cavity model, in which the ionized gas is confined to shell structures near the I-fronts, and $U$ is not calculated for the cavity interiors, where $r^2n_H\rightarrow 0$.
P11 then approximated the pressure exerted on the observed ionized gas by starlight in terms of the ionization parameter as
\begin{equation}\label{P11}
  P_{\rm stars} =  U n_H \langle h\nu \rangle \frac{L_{\rm bol}}{L_0},
\end{equation}
where $L_{\rm bol}$ and $L_0$ are the total bolometric and ionizing photon luminosity for {\it all} stars (assumed to be centered at R136) and $\langle h\nu \rangle \sim 20$ eV is average energy per ionizing photon (their equation 8). 

Although Equations \ref{P11} and \ref{P11U} can be combined and trivially reduced to Equation \ref{L11}, doing so hides the ambiguous role of radiation pressure in regions where $n_H$ vanishes. $P_{\rm stars}$ as definited by P11 represents the momentum imparted to the observed nebular gas.
L11 acknowledge this alternative definition of radiation pressure, but claim that ``it is necessary to characterize $P_{\rm dir}$ as the energy density of the radiation field, since that definition reflects the total energy and momentum budget available to drive motion.'' This definition implies that the luminosity emitted by the OB stars could impart momentum with 100\% efficiency everywhere in the nebula at once, an ideal case that could never occur in a real \hiir.
The justification brings to mind the old philosophical thought experiment about whether a tree falling in a forest makes a sound if there is no one around to hear it. If one dropped a cloud of dense, neutral gas close to the R136 star cluster, it would experience an enormous radiation pressure. But there are no dense clouds of neutral gas within the radiation-pressure-dominated region identified by L11. Instead, the interior structure of 30 Dor consists of evacuated cavities filled with hot, highly ionized plasma \cite{Townsley et al. 2006}, and pressure from either the hot, X-ray plasma or the warm, photoionized gas can exceed the radiation pressure on the cavity walls (L11, P11). Radiation pressure could have dominated the expansion of these large cavities in the past, when they were smaller (P11). It is difficult to answer this question definitively because the hot gas pressure in 30 Dor remains uncertain.

Both L11 and P11 reported values for the hot gas pressure in 30 Dor.
The pressure of the X-ray-emitting plasma can be calculated as
\begin{equation}
P_X = 1.9 n_X kT_X,
\end{equation}
where $n_X$ and $T_X$ are the density and temperature of the X-ray-emitting plasma. P11 used the results of spectral fits by Townsley et al.\ \cite{Townsley et al. 2006} while L11 performed their own spectral fitting to the same archival data plus a newer, 90-ks {\it Chandra} observation of 30 Dor \cite{Townsley 2009}.
It is difficult to derive $n_X$ accurately from the emission measure returned by spectral fitting because it depends strongly on the assumed geometry of the X-ray bubbles. L11 treated 30 Dor as a ``beach ball'' with a global, spherical geometry, which effectively minimizes $n_X$. P11 treated 30 Dor as a ``bunch of grapes,'' assuming a spherical geometry for each smaller, diffuse X-ray region identified by Townsley et al.\ \cite{Townsley et al. 2006}, which yields higher $n_X$. Spectral modeling of diffuse emission structures in regions like 30 Dor is a tremendously complicated task \cite{Townsley et al. 2011}, as any given sightline will contain plasma from multiple origins, including both stellar wind shocks and supernova, at a variety of temperatures, densities, and compositions. To illustrate the pitfalls of over-interpreting these data, I note that different approaches to the spectral fitting of the the global, diffuse X-ray emission of 30 Dor yield significantly different results. L11 fit a single-temperature plasma model and reported $kT_X = 0.64^{+0.03}_{-0.02}$ keV and absorption-corrected $L_X = 4.5\times 10^{36}$~erg~s$^{-1}$. In contrast, the most recent spectral fits by L. K. Townsley (private communication) employ 3 plasma components (plus numerous gaussian profiles to fit unidentified emission lines) ranging from 0.3--0.8 keV and yield $L_X = 1.2$--$1.9\times 10^{37}$~erg~s$^{-1}$. It is particularly difficult to discern whether regions of 30 Dor that appear X-ray dark represent the boundaries of confined, hot plasma bubbles, or whether the plasma extends behind foreground material that absorbs the soft X-rays.
The existing 114 ks combined {\it Chandra} integration represents a very shallow observation at the distance of 30 Dor when compared to observations of Galactic \hiirs.

The story of 30 Dor has a moral; given the complexity and ambiguity involved in interpreting the multiwavelength data on this {\it resolved} starburst region, investigators wishing to extend these results to draw conclusions about feedback mechanisms shaping {\it unresolved} regions at cosmological distances do so at their own risk.

\section{The Milky Way Project \hiiR\ Catalog}

Strong empirical constraints on the time-evolution of feedback-driven \hiirs\ require the comparative study of large observational samples of \hiirs. The wealth of new IR imaging data provided by the {\it Spitzer} Galactic Legacy Infrared Mid-Plane Extraordinaire (GLIMPSE) \cite{Churchwell et al. 2009} and subsequent high-resolution surveys of the Milky Way allow us to penetrate the obscuring veil of dust in the Galactic plane, revealing the structure of \hiirs\ and PDRs in unprecedented detail. Using the GLIMPSE images, Churchwell et al.\ \cite{Churchwell et al. 2006, Churchwell et al. 2007} cataloged nearly 600 IR bubbles, ring and arc-shaped structures apparent in 8~\um\ PAH emission. The majority of these structures are PDRs surrounding \hiirs, from energetic, wind-blown bubbles to low-luminosity nebulae surrounding B-type stars.

Over the past year, ${>}35,000$ internet users from around the world have been searching for bubbles in {\it Spitzer} survey images of the Galactic plane as part of the Milky Way Project (MWP; http://www.milkywayproject.org), a recent installment in the Zooniverse, the premier series of online ``citizen science'' projects (http://www.zooniverze.org). Upon creating a Zooniverse account and logging into milkywayproject.org, MWP volunteers are presented with a random image and instructed to identify and fit structures within that image that resemble bubble rims (PDRs resembling the regions in Figure~\ref{N49+M17}) with elliptical annuli (or mark the locations of bubbles that are barely resolved with boxes). By carefully combining the results from many individuals for each part of the sky, the MWP simultaneously leverages the superior pattern-recognition skills of the human eye-brain system and takes advantage of the ``wisdom of crowds.'' The MWP First Data Release presents a catalog of 5,106 \hiirs, representing an order of magnitude improvement in completeness compared to previous catalogs \cite{MWP}.

The MWP bubbles catalog will facilitate the study of triggered star formation on the Galactic scale. The idea that feedback from expanding \hiirs\ can exert external pressure on cold, molecular cloud cores, resulting in self-propagating, sequential massive star formation has been around for decades \cite{Elmegreen+Lada 1977, Sandford et al. 1982, Bertoldi 1989, Whitworth et al. 1994} but recently has seen a resurgence of observational and theoretical interest, motivated in large part by the identification of numerous instances of small bubbles, young stellar objects, masers, and other observational signposts of recent or ongoing star formation near the rims of GLIMPSE bubbles (N49 in Figure~\ref{N49+M17} provides an example, with two luminous young stellar objects and an ultracompact \hiir\ visible on the lower rim of the bubble \cite{Watson et al. 2008}). To date, most studies of triggered star formation have focused on individual, ``best-case'' regions, very round bubbles with prominent sub-clusters or smaller bubbles on their rims \cite{Deharveng et al. 2005, Zavagno et al. 2006, Zavagno et al. 2010}. In spite of the long-standing, widely popular idea that supernovae trigger star formation, far fewer candidate triggering sites have been identified near supernova remnants than near \hiirs. 

To establish that triggering merits investigation as an {\it important} mode of star formation as opposed to an astrophysical curiosity, the prevalence of triggering sites must be established statistically, and in an unbiased fashion, among representative samples of \hiirs. Both the GLIMPSE and MWP bubbles catalogs include flags for hierarchical structure, identifying smaller bubbles that could be the ``daughters'' of larger, ``parent'' bubbles. Among the MWP bubbles, 29\% are members of hierarchies \cite{MWP}. Triggered star formation need not produce observable daughter bubbles if the second generation of stars is too young to have produced \hiirs. Thompson et al.\ \cite{Thompson et al. 2011} recently found a strong correlation between young stellar objects identified as luminous mid-IR point sources and the rims of bubbles from the Churchwell et al.\ \cite{Churchwell et al. 2006} catalog. Corroboration of this correlation using the more complete sample of bubbles from the MWP would greatly strengthen the case that triggering is a prevalent mode of Galactic star formation.

\section{Summary}

In this review, I have given a brief update on recent, multiwavelength observations, particularly in the mid-IR and X-rays, that have revealed wind-blown and radiation-dominated \hiir\ structures. Observations corroborate the basic predictions of \hiir\ theory but reveal important differences, too. Theory still struggles to explain both the existence of dust within energetic \hiirs\ and its effects on \hiir\ structure. X-rays from hot, wind-shocked plasma are frequently observed to fill large, central cavities in giant \hiirs, but the X-ray luminosity is more than an order of magnitude lower than predicted by wind-blown bubble theory.

The size of the central cavities as a fraction of \hiir\ volume appears to increase with increasing ionizing luminosity (Figures \ref{N49+M17} and \ref{30Dor+W43}). This trend puts dust at larger distances from the ionizing stars in starburst regions, which reduces the effective temperature of the dust compared to static or thermal pressure-dominated \hiir\ models. Single-band IR diagnostics of extragalactic star formation rates (e.g. \cite{Calzetti et al. 2007}) have become increasingly popular in the era of large, high-resolution surveys from {\it Spitzer} and {\it Herschel}.
Because the effective dust temperature sets the shape of the IR spectral energy distribution in unresolved regions, feedback must be taken into account when calibrating the IR diagnostics.

Feedback in large, starburst regions like 30 Dor presents a particularly complicated problem.  It would be premature to assume that a single source of feedback, radiation pressure (L11), dominates such regions until we achieve a better understanding of the contributions from massive star winds and supernovae.

\hiirs\ are among the most beautiful objects appearing in astronomical images. The strong aesthetic appeal of these images helped to motivate MWP volunteers, the vast majority of whom were not professional scientists, to spend tens of thousands of person-hours finding and measuring several thousand \hiirs\ in the form of IR bubbles. The large MWP database \cite{MWP} provides an unmatched resource for statistical studies of \hiir\ evolution and star formation triggered by massive star feedback. I expect that the MWP will spawn many follow-up investigations in the coming years, involving both professional researchers and citizen scientists.

\acknowledgments I thank the organizers of the Frank N. Bash Symposium 2011 for their invitation to speak and to contribute to these proceedings. L. K. Townsley and E. W. Pellegrini provided helpful conversations and insights that greatly improved this contribution. I gratefully acknowledge support from a National Science Foundation Astronomy \& Astrophysics Postdoctoral Fellowship under award AST-0901646.


\begin{thebibliography}{99}
   \bibitem{Tielens 2005} Tielens, A.~G.~G.~M.,  {\it The Physics and 
       Chemistry of the Interstellar Medium}, Cambridge University
     Press, 2005
   \bibitem{Draine 2011} Draine, B.~T.,  {\it Physics of the Interstellar and 
       Intergalactic Medium}, Princeton University Press, 2011
   \bibitem{Stromgren 1939} Str{\"o}mgren, B.,  {\it The Physical State of 
       Interstellar Hydrogen.}, 1939, ApJ, 89, 526
  \bibitem{Spitzer 1978} Spitzer, L.,  {\it Physical processes in the 
     interstellar medium}, Wiley-Interscience, New York, 1978
  \bibitem{Castor et al. 1975} Castor, J., R.~McCray, 
    \& R.~Weaver,  {\it Interstellar bubbles}, 1975, ApJ, 200, L107
  \bibitem{Weaver et al. 1977} Weaver, R., et al.,  {\it Interstellar 
      bubbles. II - Structure and evolution}, 1977, ApJ, 218, 377 
  \bibitem{McKee et al. 1984} McKee, C.~F., D.~van Buren, 
    \& B.~Lazareff,  {\it Photoionized stellar wind bubbles in a cloudy medium}, 1984, ApJ, 278, L115 
  \bibitem{Koo+McKee 1992} Koo, B.-C.~\& C.~F.~McKee,  {\it Dynamics of wind
      bubbles and superbubbles. I - Slow winds and fast winds. II -
      Analytic theory}, 1992, ApJ, 388, 93
  \bibitem{Capriotti+Kozminski 2001} Capriotti, E.~R.~\& J.~F.~Kozminski,  {\it Relative
      Effects of Ionizing Radiation and Winds from O-Type Stars on the
      Structure and Dynamics of H II Regions}, 2001, PASP, 113, 677 
  \bibitem{Freyer et al. 2003} Freyer, T., G.~Hensler, 
    \& H.~W.~Yorke,  {\it Massive Stars and the Energy Balance of the
      Interstellar Medium. I. The Impact of an Isolated 60 M$_{\odot}$
      Star}, 2003, ApJ, 594, 888
  \bibitem{Harper-Clark+Murray 2009} Harper-Clark, E.~\& N.~Murray,
    {\it One-Dimensional Dynamical Models of the Carina Nebula
      Bubble}, 2009, ApJ, 693, 1696 
  \bibitem{Everett+Churchwell 2010} Everett, J.~E.~\& E.~Churchwell,
   {\it Dusty Wind-blown Bubbles}, 2010, ApJ, 713, 592  
  \bibitem{Pellegrini et al. 2007} Pellegrini, E.~W., et al.,  {\it A 
     Magnetically Supported Photodissociation Region in M17}, 2007, ApJ, 658, 
   1119 
  \bibitem{Krumholz+Matzner 2009} Krumholz, M.~R.~\& C.~D.~Matzner,
   {\it The Dynamics of Radiation-pressure-dominated H II Regions},
   2009, ApJ, 703, 1352  
  \bibitem{Draine 2011b} Draine, B.~T.,  {\it On Radiation Pressure in Static, 
      Dusty H II Regions}, 2011, ApJ, 732, 100 
  \bibitem{Churchwell et al. 2006} Churchwell, E., et al.,  {\it The Bubbling 
    Galactic Disk}, 2006, ApJ, 649, 759     
 \bibitem{Watson et al. 2008} Watson, C., et al.,  {\it Infrared Dust 
      Bubbles: Probing the Detailed Structure and Young Massive Stellar 
      Populations of Galactic H II Regions}, 2008, ApJ, 681, 1341 
  \bibitem{Povich et al. 2007} Povich, M.~S., et al.,  {\it A Multiwavelength 
     Study of M17: The Spectral Energy Distribution and PAH Emission Morphology 
     of a Massive Star Formation Region}, 2007, ApJ, 660, 346 
  \bibitem{Townsley et al. 2003} Townsley, L.~K., et al.,  {\it 10 MK Gas in 
     M17 and the Rosette Nebula: X-Ray Flows in Galactic H II Regions}, 2003, 
   ApJ, 593, 874 
  \bibitem{Townsley et al. 2011} Townsley, L.~K., et al.,  {\it The 
     Integrated Diffuse X-ray Emission of the Carina Nebula Compared to Other 
     Massive Star-forming Regions}, 2011, ApJS, 194, 16
  \bibitem{Pellegrini et al. 2011} Pellegrini, E.~W., J.~A.~Baldwin,
   \& G.~J.~Ferland,  
   {\it Structure and Feedback in 30 Doradus. II. Structure and Chemical Abundances}, 2011, ApJ, 738, 34 
 \bibitem{Townsley et al. 2006} Townsley, L.~K., et al.,  {\it A Chandra 
     ACIS Study of 30 Doradus. I. Superbubbles and Supernova Remnants}, 2006, 
   AJ, 131, 2140 
  \bibitem{Lopez et al. 2011} Lopez, L.~A., et al.,  {\it What Drives the 
      Expansion of Giant H II Regions?: A Study of Stellar Feedback in 30 
      Doradus}, 2011, ApJ, 731, 91
  \bibitem{Townsley 2009} Townsley, L.~K.,  {\it Diffuse X-ray Structures in 
      Massive Star-forming Regions}, 2009, AIPC, 1156, 225 
  \bibitem{Churchwell et al. 2009} Churchwell, E., et al.,  {\it The 
      Spitzer/GLIMPSE Surveys: A New View of the Milky Way}, 2009,
    PASP, 121, 213
  \bibitem{Churchwell et al. 2007} Churchwell, E., et al.,  {\it The Bubbling 
    Galactic Disk. II. The Inner $20^{\circ}$}, 2007, ApJ, 670, 428
  \bibitem{MWP} Simpson, R.~J., et al., {\it The Milky Way Project First
     Data Release: A Bubblier Galactic Disk}, 2012, MNRAS, submitted [arXiv:1201.6357]




  \bibitem{Elmegreen+Lada 1977} Elmegreen, B.~G.~\& C.~J.~Lada,  {\it
      Sequential formation of subgroups in OB associations}, 1977,
    ApJ, 214, 725
  \bibitem{Sandford et al. 1982} Sandford, M.~T., II, R.~W.~Whitaker, 
    \& R.~I.~Klein,  {\it Radiation-driven implosions in molecular clouds}, 1982, ApJ, 260, 183 
  \bibitem{Bertoldi 1989} Bertoldi, F.,  {\it The photoevaporation of 
      interstellar clouds. I - Radiation-driven implosion}, 1989, ApJ, 346, 735 
  \bibitem{Whitworth et al. 1994} Whitworth, A.~P., et al.,  {\it The 
      Preferential Formation of High-Mass Stars in Shocked Interstellar Gas 
      Layers}, 1994, MNRAS, 268, 291 
  \bibitem{Deharveng et al. 2005} Deharveng, L., A.~Zavagno, 
    \& J.~Caplan,  {\it Triggered massive-star formation on the
      borders of Galactic H II regions. I. A search for ``collect and
      collapse'' candidates}, 2005, A\&A, 433, 565
  \bibitem{Zavagno et al. 2006} Zavagno, A., et al.,  {\it Triggered 
      massive-star formation on the borders of Galactic H II regions. II. 
      Evidence for the collect and collapse process around RCW 79}, 2006, 
    A\&A, 446, 171 
  \bibitem{Zavagno et al. 2010} Zavagno, A., et al.,  {\it Star formation 
      triggered by the Galactic H II region RCW 120. First results from the 
      Herschel Space Observatory}, 2010, A\&A, 518, L81 
  \bibitem{Thompson et al. 2011} Thompson, M.~A., et al.,  {\it The 
      statistics of triggered star formation: an overdensity of massive YSOs 
      around Spitzer bubbles}, 2011, MNRAS, in press [arXiv:1111.0972]
\bibitem{Calzetti et al. 2007} Calzetti, D., et al.,  {\it The Calibration 
of Mid-Infrared Star Formation Rate Indicators}, 2007, ApJ, 666, 870 

 
\end{thebibliography}
\end{document}